\documentclass{revtex4-1}%

 \usepackage{graphics}
\usepackage{graphicx}
\usepackage{epsfig}
\usepackage{amssymb}
\usepackage{xcolor}
\usepackage{color}
\usepackage{color,soul}
\usepackage{indentfirst}
\usepackage{amsmath}
\usepackage{mathrsfs}
\usepackage[russian,english]{babel}
\begin{document}
    \title {Study of electromagnetic dipole moment, electric quadrupole moment
    and weak T-odd (CP-odd) interactions of high energy short-lived particles in
     straight crystals}
    \author{V.G. Baryshevsky }
    %\inst{Research Institute for Nuclear Problems, Belarusian State University, \\ 11 Bobruiskaya str., 220030, Minsk, Belarus}
    %\date{}
    \address{Research Institute for Nuclear Problems, Belarusian State University, \\ 11 Bobruiskaya str., 220030, Minsk, Belarus}

    \begin{abstract}{
A particle, which moves in a crystal, experiences weak
interactions with electrons and nuclei alongside with
electromagnetic interaction.
Measuring the polarization vector and the angular distribution of
charged and neutral particles scattered by axes (planes) of an
unbent (straight) crystal enables to obtain restrictions for the
EDM value and for {magnitudes} of constants describing T-odd
(CP-odd) interactions Beyond the Standard Model.
 Spin rotation and polarization conversion from vector to
tensor one and vice versa for a channelled in a crystal
$\Omega^{\pm}$ hyperon enable measuring hyperon's quadrupole
moment that is not possible to measure by use of laboratory
available noncrystalline electric fields. }
\end{abstract} %end of abstract
    %
    %\noindent \textbf{\small{Bent crystal, charm baryon, electric dipole moment, spin rotation, parity violation, magnetic
    %moment, CP violation.}}
    \maketitle

\section{Introduction}

Violation of parity (P) and time reversal (T) symmetries lead to
appearance of numerous processes allowing investigation of physics
Beyond the Standard Model. Recently, the experimental approach was
proposed \cite{bn11,b10,tau1,tau2} to search for the
electromagnetic dipole moments (EDM) of charged short-lived heavy
baryons and $\tau$-leptons using bent crystals at LHC.
{According to \cite{b29,b29a,b33} the same approach gives unique
possibility for investigation of P-{odd} T-{even} and P-{odd}
T-{odd} (CP-odd) interactions of short lived baryons
($\tau$-leptons) with electrons and nuclei.}
{Constraints on constants of the above interactions can also be
obtained}.

This paper demonstrates that measuring the polarization vector and
the angular distribution of charged and neutral particles
scattered by axes (planes) of an unbent (straight) crystal enables
to obtain restrictions for the EDM value and for {magnitudes} of
constants describing T-odd (CP-odd) interactions Beyond the
Standard Model.
It is shown that spin rotation and polarization conversion from
vector to tensor one and vice versa for a channelled in a crystal
$\Omega^{\pm}$ hyperon enable measuring hyperon's quadrupole
moment that is not possible to measure by use of laboratory
available noncrystalline electric fields.

\section{Relativistic particles spin interactions with crystals}
Since high energy particle motion in a crystal is of
quasiclassical nature, so to describe evolution of particle's spin
in electromagnetic fields inside the crystal
Thomas--Bargmann--Michel--Telegdi (T-BMT) {equations \cite{bn14}
are used.
The T-BMT equation describes} spin motion in the rest frame of the
particle, wherein spin is described by three component vector
$\vec{S}$.
In practice the T-BMT equation well describes the spin precession
in external electric and magnetic fields encountered in typical
present accelerators.
Study of the T-BMT equation enables one to determine the major
peculiarities of spin motion in an external electromagnetic field,
{ to describe the spin rotation effect for particles in a crystal
and to apply it for measuring magnetic moments of unstable
particles \cite{b10,b1,b12,b8,b6,b2,b3,b4,b5,b9}.}
However, it should be taken into account that particles in an
accelerator or a bent crystal have energy spread and move along
different orbits.
This necessitates to average the spin--dependent parameters of the
particle over phase space of the particle beam.
That is why one should always bear in mind the distinction between
beam polarization $ \vec{\xi} $ and spin vector $\vec{S}$.
Complete description of particle spin motion can be made by the
use of spin density matrices equation (in more details see
\cite{b12,b22}).
For the case of ultra relativistic baryons
{with spin $S=1/2$ } the
T-BMT equations supplied with the term, which is responsible for
interaction between particle EDM and electric field, can be
written as follows ($ \gamma \gg 1, \gamma$ is the Lorentz-factor)
\cite{b20,b16,bn11,b10}:
\begin{equation}
\frac{d \vec{\xi}}{dt}= [\vec{\xi} \times \vec{\Omega}_{magn}] + [\vec{\xi} \times \vec{\Omega}_{EDM}],
\label{eq3}
\end{equation}
where $\vec{\xi}$ is the  particle polarization vector,
$\vec{\Omega}_{magn}=-\frac{e(g-2)}{2mc} [\vec{\beta} \times
\vec{E}_{\perp}] $, $ g $ is the gyromagnetic ratio (by
definition, the particle magnetic moment
$\mu=\frac{eg\hbar}{2mc}S$, where $S$ is the particle spin),
$\vec{\Omega}_{EDM}= \frac{2ed}{\hbar}\vec{E}_{\perp}$,
$\vec{E}_{\perp}$ is an electric field component perpendicular to
the particle velocity $ \vec{v} $, the unit vector $\vec{\beta} $
is  parallel to the velocity $ \vec{v} $, quantity $D=ed$ is the
electric dipole moment.

Note that authors of \cite{tau1} use for electric dipole moment
the following expression: $\vec{\delta}=J d \mu_B \vec{s}$,
where $\mu_B=\frac{e \hbar}{2 m c}$ is the particle magneton,
{$\vec{s}$} is the spin polarization ratio, $J$ is the particle
spin, $d$ is the dimensionless factor referred to as the
gyroelectric ratio.
To avoid confusion with notation $d$, which is conventionally used
for electric dipole moment, the gyroelectric ratio is hereinafter
denoted by $d_e$.

It should be mentioned that for particles {with spin $3/2$
($\Omega^{\pm}$ hyperon)} T-BMT equations  should be supplemented
by the terms, which consider possession of electric quadrupole
moment by the particle \cite{b12,b8}.
Moreover, $\Omega^{\pm}$ hyperon could also possesses the T-odd
magnetic quadrupole moment, because its spin value is as high as
$3/2$.
%
%%%%%%%%%%%%%%%%%%%%%%%%%%%%%%%%%%%%%%%%%%%%%%%%%%%%%%%%%%%%%%%%%%%%%%%%%%%%%%%
%
Equations that compile the T-BTM one for  $\Omega^{\pm}$ hyperons
case are as follows:

\begin{eqnarray}
\frac{dS_i}{dt}= \big[\vec{S}\times\vec{\Omega_{magn}}\big]_i +
\big[\vec{S}\times\vec{\Omega_{EDM}}\big]_i +\frac{e}{3\hbar}
\varepsilon_{ikl} \varphi_{kn} \langle\hat{Q_{ln}}\rangle ,
\label{eqadd1}
\end{eqnarray}
where
$\vec{\Omega}_{magn}=-\frac{e(g-2)}{2\hbar}\lambda_c
[\vec{\beta}\times\vec{E}_{\perp}] $,
$\vec{\Omega}_{EDM}=\frac{e d_e}{2\hbar}\lambda_c\vec{E}_{\perp}$.
Here $e$ is particle electrical charge,
$\lambda_c=\frac{\hbar}{mc}$ is the Compton wavelength of a
particle, $\varphi_{kn} =\frac{\partial^2 \varphi}{\partial x_k
\partial x_n}$
is the second derivative of the electrostatic potential at the
point of particle location in crystal; $\varepsilon_{ikl}$ is the
totally antisymmetric unit tensor \cite{b1,b12},
$S_i=Sp \hat{\rho}\hat{S_i}$, ~~
$\langle\hat{Q}_{ln}\rangle=Sp \hat{\rho}\hat{Q}_{ln}$, ~~
$\hat{Q}_{ln}= \frac{3Q}{2S(2S-1)} \big\{
\hat{S_l}\hat{S_n}+\hat{S_n}\hat{S_l}-\frac{2}{3}S(S+1)\delta_{ln}\big\}
$,
$Q$ is the quadrupole moment of the particle, $\hat{\rho}$ is the
particle spin density matrix, $\hat{S}_n$ is the $n$-component of
the spin-operator $\hat{S}$.

Equation for $\langle \hat{Q}_{ln}(t) \rangle$  see in
\cite{b1,b12}.
Contribution caused by T-odd magnetic quadrupole interaction is
not included in (\ref{eqadd1})  because of its smallness, though
obtaining evaluation of this contribution is interesting.
As follows from (\ref{eqadd1}), for a particle moving in a planar
channel, formed by planes orthogonal to $x$ axis, the spin
rotation frequency caused by quadrupole moment is $\Omega_Q \simeq
\frac{eQ}{\hbar}\varphi_{xx}$. According to estimations made in
\cite{hyperon_3,hyperon_4,hyperon_5}
%
%\hl{---- см. ядерн оптику стр. 436, начало главы 16.4 ссылки на
%герштейна,Isgur and Leonard -----}
%
for quadrupole moment the value  $Q\simeq 10^{-27}$cm is expected.
 As a result, for $\varphi_{xx}\simeq 10^{18} \frac{V}{cm}^2$ we have $\Omega_Q\simeq10^{6}$s$^-1$.
 Hence distance passed by particle in crystal $L=10$cm rotation angle is $\vartheta\simeq 10^{-3}$rad,
 that corresponds to experiments for EDM limitations of heavy baryons \cite{bn11,b10,tau1,tau2}.
%\hl{-------[1,2,3,4]--------}
%
 Let us note, that in experiments in a straight crystal spin rotation caused by the magnetic moment of $\Omega^{\pm}$ hyperon is suppressed.
It is also important that for $\Omega^{\pm}$ hyperons moving in a
channel conversion of vector polarization to quadrupolarization
tensor $\langle Q_{ln} \rangle$ occurs that enables to choose the
most sensitive measuring method depending on initial conditions.

%
%%%%%%%%%%%%%%%%%%%%%%%%%%%%%%%%%%%%%%%%%%%%%%%%%%%%%%%%%%%%%%%%%%%%%%%%%%%%%%%%%%%%%%%%%%%%%%%%%%%%%%%%%%%%%%%%%%%%%%%%%%%%%%
%
According to \cite{b12,b8} investigation of spin rotation for
$\Omega ^{\pm}$-hyperons in straight and bent crystals enables to
measure the quadrupole moment of $\Omega ^{\pm}$-hyperon, which
cannot be measured by the use of available in a laboratory
noncrystalline macroscopic nonuniform electric fields.

\section{P-  and T-odd spin interactions in crystals}
General expression for the amplitude of elastic coherent
scattering of a spin $1/2$ particle by a spinless (unpolarized)
atom in presence of electromagnetic, strong and $P$-, $T$-odd weak
interactions can be written as:
\begin{equation}
\hat{F}(\vec{q})=A(\vec{q})+B(\vec{q}) \vec{\sigma}\vec{N}+B_{w}(\vec{q})\vec{\sigma}\vec{N_{w}}+B_{T}\vec{\sigma}\vec{N_{T}},
\label{eq32}
\end{equation}
where $ A(\vec{q}) $ is the spin-independent part of scattering
amplitude, which is caused by electromagnetic, strong and weak
interactions of the particle with electrons and nucleus of the
atom,
$ \hbar\vec{q}=\hbar\vec{k}^{'} - \hbar\vec{k} $ is the
transmitted momentum, $\hbar\vec{k}^{'}$ is the momentum of the
scattered particle, $  \hbar\vec{k} $ is the momentum of the
incident baryon, $\vec{k}^{'}$ and  $\vec{k}  $ are the wave
vectors,
$\vec{N}=\frac{[\vec{k}^{'}\times\vec{k}]}{[\vec{k}^{'}\times\vec{k}]}
$, $\vec{N_{w}}=\frac{\vec{k}^{'}+\vec{k}}{|\vec{k}^{'}+\vec{k}|}
$, $\vec{N_{T}}=\frac{\vec{k}^{'}-\vec{k}}{|\vec{k}^{'}-\vec{k}|}
$, $\vec \sigma = (\sigma_x,\sigma_y,\sigma_z)$ are the Pauli
matrices.

The term, which is proportional to $ \vec{\sigma}\vec{N} $, is
responsible for the contribution to scattering process, which is
caused by spin-orbit interaction.

For electromagnetic interaction {the spin-orbit interaction is
determined}  by the particle magnetic moment.
P-{odd}  T-{even} part of the scattering amplitude (it is
proportional to  $ \vec{\sigma}\vec{N_{W}} $) is determined by
P-{odd} T-{even} interactions of baryon ($\tau$-lepton) with
electrons and nuclei.
T-{odd}  part of scattering amplitude (it is proportional to  $
\vec{\sigma}\vec{N_{T}}$) is determined by the electric dipole
moment and short range particle-electron and particle-nucleus
T-{odd} interactions,
{Measurement of amplitude $B_{T}$ enables studying of physics
beyond standard model and getting limits for the constants, which
determine such interactions in hadron and lepton sectors.}

With amplitude $ \hat{F}(\vec{q}) $ one can find the cross-section
of particle scattering by a crystal and polarization vector of the
scattered particle.
{Let us now consider a thin crystal, for which effects caused by
channelling are not essential.}
The scattering cross-section for a thin crystal can be written as
\cite{b8}:

\begin{equation}
\label{eq33}
\frac{d\sigma_{cr}}{d\Omega}=\frac{d\sigma}{d\Omega}\left\{(1-e^{-\overline{u^2}
    {q^2}}) +\frac{1}{N}\left|\sum_n e^{i\vec q\vec{r}_n^0}\right|^2 e^{-\overline{u^2} {q^2}}\right\},
\end{equation}
where $\vec{r}_n^0 $ is the coordinate of the center of gravity of
the crystal  nucleus, $\overline{u^2}$ is the mean square of
thermal oscillations of nuclei in the crystal. The first term
describes incoherent scattering,
{ caused by the thermal vibration of {crystal nuclei} and the
second one describes the coherent scattering due to periodic
arrangement of crystal nuclei (atoms).}

Quantity $ \frac{d\sigma}{d\Omega} $  describes cross-section of
baryon scattering
 by atoms of the crystal:
\begin{equation}
\label{eq34} \frac{d\sigma}{d\Omega}= tr \hat{\rho}
\hat{F^{+}}(\vec{q})\hat{F}(\vec{q}),
\end{equation}

\noindent where $ \hat{\rho} $ is the spin density matrix of the
incident particle.

The polarization vector of the particle, which has undergone a
single scattering event, can be
found using the following expression: %\cite{b27}
\begin{equation}
\label{eq35} \vec\xi= \frac{\mbox{tr}\hat{\rho}
\hat{F^+}\vec\sigma \hat{F}}{\mbox{tr}\hat{\rho} \hat{F^+}
\hat{F}} = \frac{\mbox{tr}\hat{\rho} \hat{F^+} \vec\sigma
\hat{F}}{\frac{d\sigma}{d\Omega}}.
\end{equation}

\noindent Using (\ref{eq32})
 one can obtain the following expressions for
polarization vector of the scattered particle \cite{b33}:
\begin{equation}
\label{eq36}
\vec{\xi}=\vec{\xi_{so}}+\vec{\xi_{w}}+\vec{\xi_{T}},
\end{equation}
where $ \vec{\xi_{so}} $ is the contribution to polarization
vector due to spin-orbit interaction, $ \vec{\xi_{w}} $ is that
due to weak parity violating interaction, $ \vec{\xi_{T}} $ is
contribution caused by $T$-odd interaction:

\begin{eqnarray}
\label{eq371} \vec{\xi}_{so} & = & \left\{(|{A}|^2 - |B|^2)
\vec\xi_0 + 2
|B|^2 \vec N (\vec N \cdot\vec\xi_0)+ \right. \nonumber \\
&  & \left.  +  2 {Im} ({A}B^*)[\vec N \times \vec\xi_0] +2 \vec N {Re} ({A}B^*)\right\}\cdot \left(\frac{d\sigma}{d\Omega}\right)^{-1}, \nonumber \\
\end{eqnarray}
\begin{eqnarray}
\label{eq372} \vec{\xi}_{w}  & = &  \left\{(|{A}|^2 - |B_{w}|^2)
\vec\xi_0 + 2
|B_{w}|^2 \vec{N_{w}} (\vec{N_{w}} \cdot \vec\xi_0)+ \right. \nonumber \\
&  & \left.  + 2 {Im} ({A}B^*_{w})[\vec{N_{w}} \times \vec\xi_0]+2
\vec{N_{w}} {Re} ({A}B^*_{w})\right\}\cdot
\left(\frac{d\sigma}{d\Omega}\right)^{-1}, \nonumber \\
\end{eqnarray}
\begin{eqnarray}
\label{eq373} \vec{\xi}_{T} & = & \left\{(|{A}|^2 - |B_{T}|^2)
\vec\xi_0 + 2
|B_{T}|^2 \vec{N_{T}}  (\vec{N_{T}} \cdot \vec\xi_0)+ \right. \nonumber \\
&  & \left.  +  2 {Im} ({A}B^*_{T})[\vec{N_{T}} \times
\vec\xi_0]+2 \vec{N_{T}}  {Re} ({A}B^*_{T})\right\}\cdot
\left(\frac{d\sigma}{d\Omega}\right)^{-1},
\end{eqnarray}
where $\vec\xi_0$ is the polarization vector of a particle
incident on a target

\noindent The differential cross-section in the same case reads as
follows:
\begin{eqnarray}
\label{eq381}
& & \frac{d\sigma}{d\Omega}=\mbox{tr}\rho F^+ F = \nonumber \\
& &  = |{A}|^2 + |B|^2+|B_{w}|^2+|B_{T}|^2 +
2Re({A}B^*)\vec N \cdot \vec\xi_0 + \nonumber \\
& & + 2Re({A}B_{w}^*)\vec{N_{w}} \cdot \vec\xi_0+
2Re({A}B_{T}^*)\vec {N_{T}} \cdot \vec\xi_0.
\end{eqnarray}

While deriving expressions (\ref{eq371})-(\ref{eq373}) and
(\ref{eq381}) the small terms containing productions  $ B B_{T} $,
$B B_{w}  $ and $ B_{w} B_{T} $,
which describe interference between spin-orbit P-odd T-even and
P-odd T-odd interactions, are omitted.

{These terms are much smaller as compared to those ones
proportional to productions $AB_{w}$ and $AB_{T}$, which describe
interference of weak interaction with strong and electromagnetic
interactions}.

However, the omitted here contributions could be significant for
neutral particles (see comments hereinafter).
%-----------------------------------------------------------------------------------------------

In case of neutral particles there is no Coulomb scattering,
therefore, the terms proportional to $B B_w$ and $B B_T$ could
also significantly contribute to anisotropy and spin rotation.
In this case expression {(\ref{eq372})} for $\xi_w$ should be
appended with addition as follows:
\begin{equation}
\Delta \xi_w = \left\{2 Re(B^* B_w) \left[ (\vec{\xi_0} \vec{N})
\vec{N}_w  + (\vec{\xi_0} \vec{N}_w) \vec{N}\right] + 2 Im (B^*
B_w)[\vec{N} \times \vec{N}_w] \right\} \frac{d \sigma}{d
\Omega}^{-1} \,. \label{eq:addBw}
\end{equation}
Expression {(\ref{eq373})} for $\xi_T$ should be appended with the
following summand:
\begin{equation}
\Delta \xi_T = \left\{2 Re(B^* B_T) \left[ (\vec{\xi_0} \vec{N})
\vec{N}_T  + (\vec{\xi_0} \vec{N}_T) \vec{N}\right] + 2 Im (B^*
B_T)[\vec{N} \times \vec{N}_T] \right\} \frac{d \sigma}{d
\Omega}^{-1} \,. \label{eq:addBT}
\end{equation}
Expression {(\ref{eq381})} for $\frac{d \sigma}{d \Omega}$ should
be appended with summand  $\frac{d \sigma_{app}}{d \Omega}$ as
follows:
\begin{equation}
\frac{d \sigma_{app}}{d \Omega} = -2 Im (B^* B_w) \vec{\xi_0}
[\vec{N} \times \vec{N}_w] - 2 Im (B^* B_T)\vec{\xi_0} [\vec{N}
\times \vec{N}_T] . \label{eq:add.sigma}
\end{equation}

%-----------------------------------------------------------------------------------------------
\noindent According to (\ref{eq371}) the angle of polarization
vector rotation for a baryon scattered in a crystal is determined
by rotations around three mutually orthogonal directions (see
terms proportional to $ N$, $N_{w}$, $N_T $).
The indicated rotations are determined by electromagnetic, strong and weak  P, T-{odd}  interactions.
It should also be noted that initially unpolarized particle beam
($\xi_0=0 $) in a crystal acquires polarization
directed along one of three vectors $\vec N$, $\vec{N_{w}}$, $\vec
{N_{T}}$, which carries information about all types of
interaction too.
According to (\ref{eq381}) amplitudes interference results in
asymmetry in scattering caused by orientation of vectors $\vec
{N_{T}}$, $\vec N$, $\vec{N_{w}}$ with respect to $\vec\xi_0,
\vec{k}'$ and $ \vec{k}$ .
Therefore, the angular distribution of scattered particles
intensity is anisotropic.
Thus, measurements of the rotation angle and of the angular distribution of intensity for a
particle beam scattered by crystal axes enables to study T-{odd }
interactions of positive(negative) charged and neutral short-lived
baryons and $\tau$-leptons.
In particular, such measurements allow one to obtain restrictions
on electric dipole moment of short-lived particles and { other
T-odd interactions in hadron and lepton sectors.
According to {\cite{b29,b29a,b33}}
 the mentioned interactions can be much stronger
than those predicted by Standard Model. Obtaining experimental
restrictions on this interactions is important
{\cite{b28,CP1,CP2}}.
%

%This can be realized, for example, by using set of crystals with
%axes directed at small angle with respect to momentum of the
%scattered particles and detection of either nuclear reaction or
%ionizing losses inside crystal detector.
%
Computer modelling is essential for further analysis.
{ Note that analyzing angle of rotation and angular distributions
one should consider
trajectories of the scattered particles with azimuth angles, which
are are in the vicinity  $\varphi$ and $\varphi + \pi$ ($z$ axes
is directed along the momentum of the incident particle). For such
particles contributions to spin rotation caused by EDM (T-odd
interaction) have opposite signs.}
As a result the T-{odd} spin rotation can be observed in unbent crystal if we use subtraction  of the measurements  results for angle ranges $
\varphi $ and $ \varphi + \pi $ from each other.
Such procedure leads to summation of contributions from T-odd
rotation.
Simultaneous measurement of spin orientation for all $ \varphi $
values (as well as for all polar angles) provides intensity
increase.

Let us now evaluate the described effects starting from estimation
of anisotropy of the angular distribution of scattered particles.
According to (\ref{eq381}) the anisotropy value is determined by
interference of amplitude $A$ with amplitudes $B$, $B_w$ and
$B_T$.
The respective contributions to the intensity of scattered
particles associated with amplitudes' interference are given by
the following ratios:
\begin{eqnarray}
\label{eq:add1}
& G=\frac{2 Re(AB)}{|A|^2}\,, \nonumber \\
& G_w=\frac{2 Re(AB_w^*)}{|A|^2} \,, \\
& G_T=\frac{2 Re(AB_T^*)}{|A|^2} \,.\nonumber
\end{eqnarray}

To observe anisotropy $G$, $G_w$ or $G_T$ the relative  value of
fluctuations in number of scattered particles $\delta \approx
\frac{1}{\sqrt{N}}$ should be made smaller as compared to $G$,
$G_w$ and $G_T$, respectively.
In other words the number of scattered particles should satisfy
the conditions as follows:
\begin{equation}
N > \frac{1}{G^2}\,; \,\frac{1}{G_w^2}\,; \, \frac{1}{G_T^2} \, .
\label{eq:add2}
\end{equation}
%%%%%%%%%%%%%%%%%%%%%%%%%%%%%%%%%%%%%%%%%%%%%%%%%%%%%%%%%%%

Let us now consider expressions (\ref{eq:add1}) and
(\ref{eq:add2}) in more details.
Amplitude $A$ is the sum of amplitudes $A_{coul}$ and $A_s$ caused
by Coulomb and strong nuclear scattering, respectively.
Amplitude of spin-orbit scattering $B=B_{magn}+B_{so}$  is caused
by particle magnetic moment interaction $B_{magn}$ and by strong
nuclear spin-orbit interaction $B_{so}$.
Let us start with evaluation of anisotropy $G$ caused by magnetic
and strong nuclear spin-orbit interaction:
\begin{equation}
G=\frac{2 Re (A_s B_{magn}^* + A_s B_{so}^* +A_{coul} B_{magn}^* +
A_{coul} B_{so}^*)}{|A_{coul}+A_s|^2}. \label{eq:add3}
\end{equation}
For further analysis let us pay attention to the fact that from
results presented in \cite{Luboshits} the amplitude of scattering
of a particle, which possesses magnetic moment, by a Coulomb field
can be expressed as follows:
\begin{equation}
A=A_{coul}(\vartheta)+i\frac{1}{2} \left( \frac{g-2}{g} \,
\frac{\gamma^2-1}{\gamma} + \frac{\gamma -1}{\gamma} \right)
\vartheta A_{coul}(\vartheta) \vec{\sigma} \vec{N}\,.
 \label{eq:add4}
\end{equation}

\noindent From (\ref{eq:add4}) the following expression for
amplitude $B_{magn}$ can be obtained:
\begin{equation}
B_{magn}=i\frac{1}{2} \left( \frac{g-2}{g} \,
\frac{\gamma^2-1}{\gamma} + \frac{\gamma -1}{\gamma} \right)
\vartheta A_{coul}(\vartheta) \,.
 \label{eq:add6}
\end{equation}

\noindent As a result in case of elastic Coulomb scattering $Re
(A_{coul} B_{magn}^*) = 0$.
Analysis shows that for the contribution to the scattering
amplitude caused by the spin-orbit strong interaction one can
obtain the expression similar to (\ref{eq:add6}) by using the
optical model of a nucleus:
\begin{equation}
B_{so}=i\frac{1}{2} \left( \frac{g_{so}-2}{g} \,
\frac{\gamma^2-1}{\gamma} + \frac{\gamma -1}{\gamma} \right)
\vartheta A_{s}(\vartheta) \,.
 \label{eq:add7}
\end{equation}
Introduced in (\ref{eq:add7}) quantity $g_{so}$ is similar to
magnetic $g$-factor and depends on the particle energy.

\noindent From (\ref{eq:add7}) it follows that $Re (A_s
B_{so}^*)=0$. Therefore,
\[
Re(A B^*) \simeq A_s^{\prime \prime} A_{coul}^{\prime} \left(
\frac{g-2}{4} +  \frac{g_{so}-2}{4} \right) \gamma \vartheta,
\]
here $A_{coul}^{\prime} = Re A_{coul}$ and $A_s^{\prime \prime} =
Im A_s$).
The above expression is obtained with consideration of the
imaginary part of Coulomb amplitude $A_{coul}^{\prime \prime}$ to
be smaller as compared to its real part $A_{coul}^{ \prime}$:
$A_{coul}^{\prime \prime}$ is $Z \alpha$ times smaller as compared
to $A_{coul}^{ \prime}$ (here $\alpha$ is the fine-structure
constant).

\noindent To evaluate the imaginary part of amplitude $A_s^{\prime
\prime}$ of baryon scattering by a nucleus let us use a
model of diffraction scattering.
As a result in  eikonal approximation $A_s^{\prime \prime}$ reads
as follows:
\begin{equation}
A_s^{\prime \prime} = R_{nuc} \frac{J_1(R_{nuc} k
\vartheta)}{\vartheta},
 \label{eq:add9}
\end{equation}
where $R_{nuc}$ is the radius of nucleus, $J_1$ is the Bessel
function of the first order.
From (\ref{eq:add9}) it follows that for scattering angles
$\vartheta \leq \frac{1}{kR_{nuc}}$ the imaginary part of
scattering amplitude $A_s^{\prime \prime} \approx  k R_{nuc}^2$.
The real part of Coulomb amplitude $A_{coul}^{\prime} = \frac{Z
\alpha}{k \vartheta^2}$ becomes comparable or even greater than
$A_s^{\prime \prime}$ for scattering angles $\vartheta \leq
\frac{\sqrt{Z \alpha}}{k R_{nuc}}$.

For a baryon with energy 1~TeV the scattering angle is $\vartheta
\leq \sqrt{Z \alpha} \cdot 10^{-5}$.
Therefore, in the range of angles, within which $A_s^{\prime
\prime} \simeq A_{coul}^{\prime}$, parameter $G \simeq \left(
\frac{g+g_{so}}{2} -2 \right) \gamma \vartheta \simeq \left(
\frac{g+g_{so}}{2} -2 \right) 10^{-2}$. As a result, to comply
inequality  $N > \frac{1}{G^2}$, the number of scattered particles
should be $N \simeq 10^4 \div 10^5 $.

Parameter $g_{so}$ depends on the energy of the incident particle
in contrast to $g$-factor, which does not at currently present
particle energies. This fact makes it possible to distinguish
contributions from $g$ and $g_{so}$ from each other.

Number $N_{\Lambda_c^+}$ of charmed lambda baryons produced by
$10^{17} \div 10^{18}$ photons in a tungsten target can be find
using data published in \cite{bn11} that gives $N_{\Lambda_c^+}
\approx 10^{13} \div 10^{14}$.
These particles, move within angle $\frac{1}{\gamma} \approx
10^{-3}$.

Hereinafter let us consider the incidence of $\Lambda_c^+$ baryons
on a target at the angle, which is equal or greater as compared to
the Lindhard critical angle, which is for Si (Ge) of order
$\vartheta_L \approx 6(7) \cdot 10^{-6}$\,rad.
Let us consider angles range $\Delta \vartheta \sim 10^{-5}$,
which amounts several Lindhard angles. Within this range
$\frac{\Delta \vartheta}{\gamma} \sim 10^{-2}$, therefore, the
number
 of $\Lambda_c^+$ baryons within this range is $\Delta N_{\Lambda_c^+} \approx 10^{11} \div
 10^{12}$.
 Let us now give evaluation for number of these particles
 scattered in crystal of thickness $l=0.1$\,cm:
 \[
 N \approx \frac{d \sigma}{d \Omega} \Delta \Omega \, N_{at } \,
 l \,
 \Delta N_{\Lambda_c^+} \approx 10^{-2} \Delta N_{\Lambda_c^+}
 \approx 10^9 \div 10^{10},
 \]
here $N_{at}$ is the number of atoms in 1cm$^3$ of target.
This value complies condition $N > \frac{1}{G^2} \approx 10^4 \div
10^5$ that enables carrying measurement of baryon magnetic moment
by analysis of anisotropy of angular distribution of scattered
baryons.

Let us now evaluate baryon scattering anisotropy, which is caused
by T-odd processes, for example, presence of particle EDM of order
$e \cdot d \sim 10^{-17}$\,e$\cdot$cm.

For a particle possessing EDM, which is scattered in an electric
field, the following expression for amplitude $B_T$ in eikonal
approximation can be used:
\begin{equation}
B_T(\vartheta)=i d \, k \, \vartheta A_{coul}=\frac{d}{\lambda_c}
\gamma \vartheta A_{coul}(\vartheta) \label{eq:add10}
\end{equation}
where $ \lambda_c=\frac{\hbar}{mc}$ is the Compton wavelength of
the particle.
For the supposed EDM value amplitude $B_T$ appears to be three
orders smaller as compared to amplitude of magnetic scattering
$B_{magn}$.
Therefore, $G_T \approx 10^{-3} G$ and the number of particles
required to make anisotropy, which is associated with amplitude
$B_T$, is higher $10^6$ times i.e. required number of particles $N
> 10^{10} \div 10^{11}$.
Recall that hereinabove for the target with high $Z$ and thickness
$l=0.1$\,cm we have obtained $N \approx 10^9 \div 10^{10}$.
Therefore, increasing the target thickness to $l=1$\,cm (such
target is still quite thin) and optimizing all the experiment
parameters one could expect to observe EDM-caused anisotropy and
that caused by other T-odd interactions. Such possibility is
important for studying EDM and T-odd interactions of short-lived
particles.

Let us now dwell on possibility to investigate EDM and other T-odd
interactions for $\tau$-leptons, for which do not undergo strong
interactions.
The EDM-caused anisotropy for $\tau$-leptons is suppressed,
because the contribution to the cross-section, which is caused by
interference of Coulomb amplitude and amplitude $B_T$ defined by
(\ref{eq:add10}), is equal to zero. The non-zero summand is due to
interference of Coulomb amplitude and T-odd amplitude caused by
neutral currents.
Restriction for the magnitude of the latter amplitude enables to
evaluate the constant of corresponding interaction.
The mentioned interaction is now persistently studied for
electrons in optical experiments with atoms
\cite{CP2}.

Let us now evaluate the angle of spin rotation for a scattered
particle, which possesses EDM.
According to (\ref{eq373}) the additional polarization component,
which arises due to rotation around $\vec{N}_T$ is determined by
the expression as follows:
\begin{equation}
\Delta \xi_{T\,rot}=\frac{2 Im (A B_T^*)}{|A|^2} .
\label{eq:add11}
\end{equation}
Therefore, for a particle scattered in Coulomb field the angle of
spin rotation can be evaluated using (\ref{eq371}) as follows:
\begin{equation}
\vartheta_s \simeq \Delta \xi_{T\,rot} \simeq \frac{d}{\lambda_c}
\gamma \vartheta. \label{eq:add12}
\end{equation}
According to (\ref{eq:add12}) the angle of spin rotation grows
with growth of scattering angle and $\gamma$ (with scattering
angle growth the angles of spin rotation caused by other T-odd
interactions also grow).
For $d=10^{-17}$~cm and $\lambda_c=10^{-14}$~cm the angle of
rotation is $\Delta \xi_{T\,rot} \simeq 10^{-3} \gamma \vartheta$,
therefore for a particle with $\gamma \sim 10^3$ the angle of spin
rotation is expected to be as high as the scattering angle:
$\vartheta_s \simeq \Delta \xi_{T\,rot} \simeq \vartheta$.
Hence, for $\gamma = 10^3$ and scattering angle $\vartheta \simeq
10^{-4} \div 10^{-5}$ the angle of spin rotation $\vartheta_s
\simeq \Delta \xi_{T\,rot} \simeq 10^{-4} \div 10^{-5}$.
The number of detected particles is $N > \frac{1}{\vartheta_s^2}
\simeq 10^8 \div 10^{10}$.
Recall that in this case spin rotates around direction
$\vec{N}_T$, which is determined by the direction of transferred
momentum $\vec{q}$.

Investigation of anisotropy and rotation angle caused by amplitude
of of weak P-odd interaction $B_w$, which is determined by neutral
P-odd currents
is of interest for short-lived baryons and $\tau$-leptons.
Recall that for electrons the neutral currents were observed in
two types of experiments: at deep inelastic scattering at SLAC
accelerator \cite{rins_85,b26+,SLAC1,SLAC2} and in optical
experiments in Novosibirsk \cite{Barkov}.

\section{Conclusion}
 The channelled particle, which moves in a crystal, besides
electromagnetic interaction experiences weak interaction with
electrons and nuclei, as well as strong interaction with nuclei.
Measurements of polarization vector and angular distribution of
charged and neutral particles particles scattered by axes (planes)
of unbent crystal enable to obtain limits for the EDM value and
for the values of constants describing P- and T-odd interactions
beyond Standard Model.
The experimental capabilities available at LHC make it possible
for a channelled in a crystal $\Omega^{\pm}$ hyperon to observe
and apply spin rotation effect and polarization conversion from
vector to tensor one and vice versa  for measuring hyperon's
quadrupole moment that is not possible to measure by use of
laboratory available noncrystalline electric fields.

 \end{document}